\documentclass{emulateapj}

\slugcomment{To appear in ApJ Letters}

\makeatletter

\newenvironment{inlinefigure}{%
\def\@captype{figure}%
\noindent\begin{minipage}{0.999\linewidth}\begin{center}}
{\end{center}\end{minipage}\smallskip}
\makeatother

\newcommand{\HII}{H{\sc ii} }

\newcommand{\HeIII}{He{\sc iii} }

\newcommand{\simgt}{\lower.5ex\hbox{$\; \buildrel > \over \sim \;$}}

\begin{document}

\title{Formation of Massive Primordial Stars in a Reionized Gas}

\author{Naoki Yoshida\altaffilmark{1}, Kazuyuki Omukai\altaffilmark{2}, Lars Hernquist\altaffilmark{3}}
\altaffiltext{1}{Department of Physics, Nagoya University, Furocho, Nagoya, Aichi 464-8602, Japan}
\altaffiltext{2}{National Astronomical Observatory of Japan, 2-21-1 Osawa, Mitaka, Tokyo 181-8588, Japan}
\altaffiltext{3}{Harvard-Smithsonian Center for Astrophysics, 60 Garden Street, Cambridge, MA02138}

\begin{abstract}
We use cosmological hydrodynamic simulations with
unprecedented resolution to
study the formation of primordial stars in an ionized gas at high redshifts.  
Our approach includes all the relevant atomic and molecular physics to follow the
thermal evolution of a prestellar gas cloud to very high densities of 
$\sim 10^{18} {\rm cm}^{-3}$. 
We locate a star-forming gas cloud within a reionized region in our cosmological simulation.
The gas cloud cools by HD line cooling down to a few tens Kelvin,
which is lower than possible by H$_2$ cooling only.
Owing to the low temperature, the first run-away collapse
is triggered when the gas cloud's mass is $\sim 40 M_{\odot}$.
We show that the cloud core remains stable
against chemo-thermal instability and also against gravitational deformation
throughout its evolution. Consequently, a single proto-stellar seed is formed,
which accretes the surrounding hot gas
at the rate $\dot{M}\simgt 10^{-3} M_{\odot}/{\rm yr}$.
We carry out proto-stellar evolution
calculations using the inferred accretion rate. 
The resulting mass of the star when it reaches the zero-age main 
sequence is $M_{\rm ZAMS}\sim 40 M_{\odot}$. We argue that, since the obtained
$M_{\rm ZAMS}$ is as large as the mass of the collapsing parent cloud, 
the final stellar mass should be close to this value. 
Such massive, rather than exceptionally 
massive, primordial stars are expected to cause
early chemical enrichment of the Universe by exploding as black hole-forming 
super/hypernovae, and may also be progenitors of high redshift $\gamma$-ray bursts. 
The elemental abundance patterns of recently discovered hyper 
metal-poor stars suggest that they might have been born from the 
interstellar medium that was metal-enriched by supernovae of these 
massive primordial stars.
\end{abstract}


\section{Introduction}

Recent observations of high redshift galaxies (Kashikawa et al. 2006),
quasars (Fan et al. 2006), and gamma-ray bursts (Totani et al. 2006)
provide rich information on the ionization structure of the
intergalactic medium (IGM) at $z\sim 6-7$. The observations indicate
that a large fraction of the IGM was already ionized when the Universe
was less than one billion years old, but suggest also that a
substantial fraction was still neutral. Combined with
polarization measurements
of the cosmic microwave background (CMB) radiation by
the WMAP satellite (Spergel et al. 2007; Page et al. 2007), it is
inferred that cosmic reionization was extended and lasted for several
hundred million years (see, e.g. McQuinn et al. 2006).

The standard cosmological model based on cold dark matter and dark
energy predicts that the first stars form in low-mass halos at
high redshifts (Couchman \& Rees 1986; Yoshida et al. 2003).
Reionization subsequently proceeds in a patchy manner, whereby \HII
regions develop around individual sources and eventually percolate
(Gnedin \& Ostriker 1997; Miralda-Escude et al. 2000;
Sokasian et al. 2003, 2004; Furlanetto et al. 2004, 2006; Zahn et
al. 2007).  Since the sources, massive stars, are
expected to emit ionizing photons for only 
a few to ten million years, the reionized IGM will start recombining 
and cooling when the stars die.
We study the formation of primordial stars in such reionized regions.

While there have been a number of numerical studies of the formation
of primordial stars in a neutral gas (Omukai \& Nishi 1998; Abel et
al. 2002; Bromm et al. 2002; Yoshida et al. 2006, hereafter Y06), the
evolution of a primordial gas cloud during/after reionization has not
been examined in detail.  It is well-known that the thermal evolution
of a recombining gas differs significantly from initially
neutral primordial gas (e.g., Nagakura \& Omukai 2005; Johnson \& Bromm 2006). 
Moreover, 
it has been suggested that cooling by HD 
molecules leads to the formation of low-mass, possibly near solar-mass,
primordial stars (Uehara \& Inutsuka 2000; Nakamura \& Umemura 2002).
It is important to
determine whether or not such small mass gas clumps are formed in a
reionized IGM at high redshifts.

Earlier (Yoshida et al. 2007; hereafter Y07), we used a
large cosmological simulation to study the evolution of
early relic \HII regions until second-generation gas clouds are
form.  In this {\it Letter}, we further explore the evolution of 
these prestellar gas clouds. We investigate in detail the structure of 
such gas clouds and examine whether they become chemo-thermally
unstable to fragmentation into smaller clumps.  We then compute the
gas mass accretion rate and use it as an input to a proto-stellar
calculation.

\section{Physical conditions around the protostar}
We use the simulation outputs of Y07. Briefly, they carried out 
radiation-hydrodynamics calculations of the formation and evolution of 
early \HII/\HeIII regions around 
the first stars.  
They showed that the ionized gas cools and re-collapses
in a growing dark matter halo within a hundred million years,
to form so-called second generation star(s). They also examined
the impact of an external far-UV background radiation field.

\begin{inlinefigure}
\resizebox{9.2cm}{!}{\includegraphics{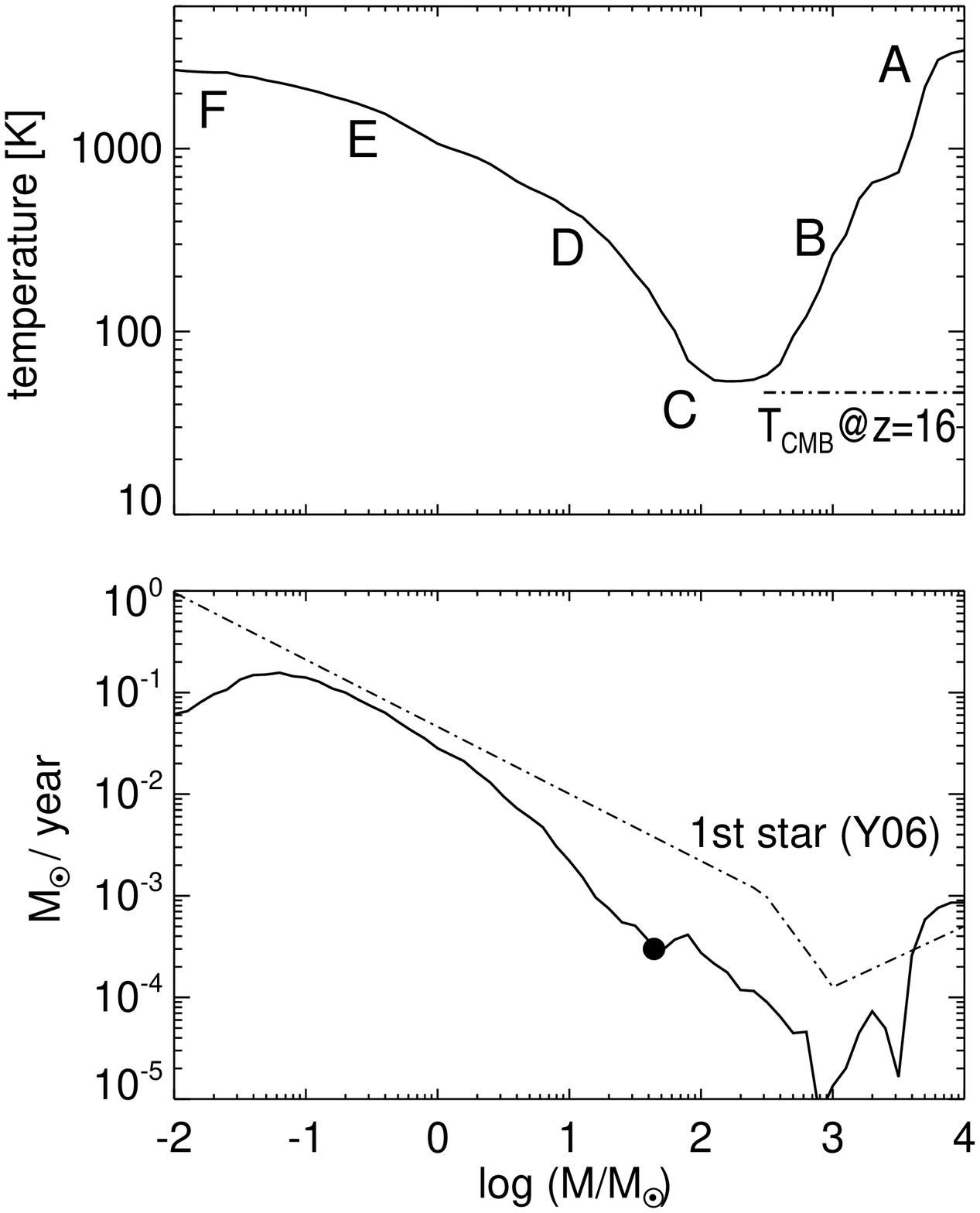}}
\caption{Top: Radial temperature profile as a function of enclosed gas mass.
We explain the characteristic features as follows:
(A) Efficient molecular hydrogen production in the recombining gas 
enables rapid cooling to a few hundred Kelvin;
(B) HD molecules are formed and the gas cools further below 100 K;
(C) heating by the CMB
floors the gas temperature, and the cloud starts to contract
gravitationally; (D) three-body reactions kick in and the core becomes fully molecular;
(E) the core is already optically thick to H$_2$ lines but cools by collision-induced emission;
(F) the core becomes optically thick to continuum.
Bottom: Instantaneous gas mass accretion
rate as a function of enclosed gas mass. The solid circle
indicates where the enclosed gas mass exceeds the locally
estimated Bonnor-Ebert mass.
The dot-dashed
line is the accretion rate around the first generation star 
in the simulation of Y06.
}
\end{inlinefigure}

\noindent We use the output of the fiducial run of Y07, in which a
second generation star forms at $z=16$ in a halo with mass
$\sim 10^7 M_{\odot}$, whose virial temperature is $\sim 10^4$ K.
In addition to the atomic and molecular physics described in Y07,
we have also implemented computation of local optical depth 
to continuum radiation using the Planck opacity table of Lenzuni et al. (1991).
The continuum opacity is used to evaluate the net cooling rate
by collision-induced emission.  

Figure~1 shows the radial profiles of gas temperature and 
gas mass accretion rate in the star-forming gas cloud. We use
the final output of the simulation when the central
density is $10^{18} {\rm cm}^{-3}$. 
The density profile at this time is 
close to a single power-law of $\rho \propto r^{-2.3}$ over
nearly 20 decades in density.
The characteristic features of the temperature profile can be understood
by appealing to various atomic and molecular processes,
as denoted in the figure 
and explained in the caption.
Radiative cooling of a molecular gas is 
halted by effective {\rm heating} from CMB photons
that have a temperature of $T_{\rm CMB} = 2.728 (1+z)$ at redshift $z$.
The minimum temperature of the molecular gas cloud is limited
to this value. 
We calculate the Bonnor-Ebert mass using the density, temperature
and molecular fraction profiles; the cloud
is dynamically unstable at a mass scale of $\sim 40 M_{\odot}$.
This is the mass for the first run-away collapse of the 
star-forming cloud.  Note that it is essentially the Jeans mass
at the temperature minimum.

The gas mass accretion rate shown in the bottom panel of Figure~1
reflects the temperature structure. 
Note that, overall, the accretion rate is much smaller than that for
the first star simulated by Y06 (dot-dashed line).
This can be understood by the fact that the mass accretion rate is
roughly $\dot{M} \propto c_{\rm s}^3/G$, where $c_{\rm s}$
is the sound speed. HD cooling enables the gas cloud to cool
to lower temperatures than is possible by H$_2$ cooling,
and hence the accretion rate within the cloud is smaller. 

We do not consider cooling by heavy elements in our simulation, 
assuming that the gas is not chemically enriched.  However,
the overall thermal evolution will not differ much from 
that shown in Figure~1, unless a significant amount of dust
is present, because cooling by metal lines 
is efficient only at low densities. HD cooling alone can 
bring the gas to $T_{\rm CMB}$, and the
gas temperature cannot fall below this even with atomic metal 
cooling.

\begin{inlinefigure}
\resizebox{9.3cm}{!}{\includegraphics{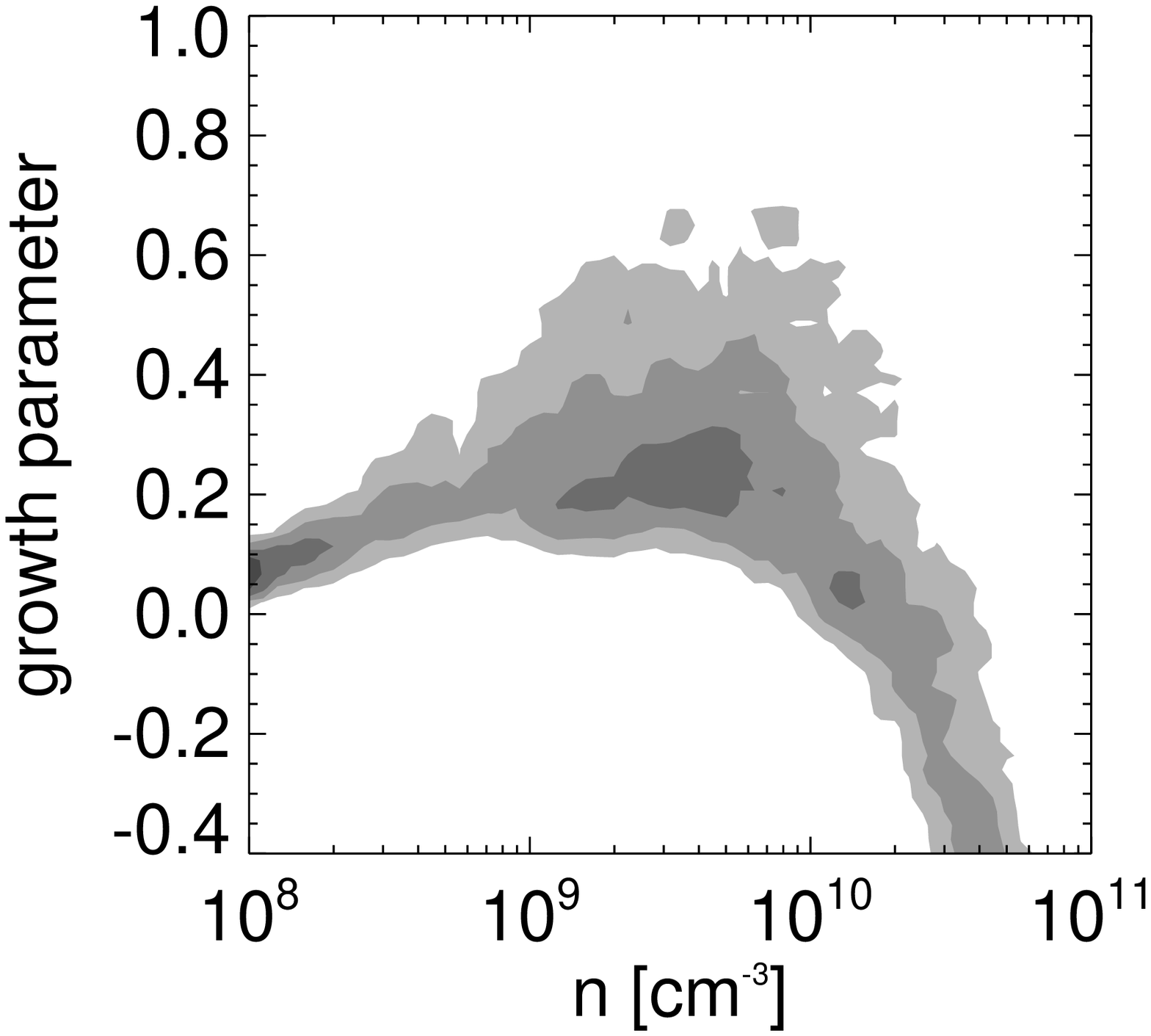}}
\caption{Growth parameter $Q=t_{\rm dyn}/t_{\rm g}$
versus gas density. We use an output when the central
density is $10^{12} {\rm cm}^{-3}$ and show the distribution of the
gas elements in this plane as grey-contours. All of the gas elements are
in the stable region $Q<1$, where the dynamical time is shorter that
the growth time of the perturbations.
\label{fig:instability}}
\end{inlinefigure}

\section{Thermal instability and fragmentation}
An important question is whether or not the gas cloud
becomes thermally unstable owing to efficient molecule production
and cooling which operates at densities $ n > 10^{8} {\rm cm}^{-3}$. 
We calculate the growth rate of isobaric perturbations
following Omukai \& Yoshii (2003).
We solve the dispersion relation
for perturbations in temperature, density, and molecular fraction,
$(\delta T, \delta \rho, \delta f) \propto \exp (\omega t)$, 
under isobaric conditions.
We define the growth parameter as the ratio
of the characteristic growth time scale $t_{\rm g}$ of the perturbations to the dynamical time
$t_{\rm dyn}$:
\begin{equation}
Q = \omega t_{\rm dyn} = \frac{t_{\rm dyn}}{t_{\rm g}}.
\end{equation}
Figure~2 shows $Q$ as a function of local gas density.
Clearly, all the gas elements are in the stable region
$Q < 1$,
where the dynamical time is shorter than the growth time scale.
We thus conclude that the chemo-thermal instability does not
operate in this regime. 
Note that, because of the harder effective equation of state
than the initially neutral primordial gas
in the relevant density and temperature range, 
gravitational deformation is less effective for the
gas cloud. We followed Omukai et al. (2005) and checked that 
the cloud is also stable against gravitational deformation.
Our simulation indeed shows that the 
cloud core remains stable against fragmentation, as expected.

\section{Proto-stellar evolution}
We carry out proto-stellar evolution calculations to study
the growth of the protostar in detail.
We employ the scheme
described by Stahler, Shu, \& Taam (1980) as modified by 
Stahler et al. (1986) and Omukai \& Palla (2003).
We use the mass accretion rate shown in Figure~1 as input to the protostellar 
evolution code, assuming that the gas is accreted in an approximately 
spherical manner. 
As is shown in Figure~1, the central part
is accreting the surrounding gas at a rate $\sim 10^{-3} M_{\odot} / {\rm yr}$
and thus a star with mass $\sim 10 M_{\odot}$ will form
within $10^{4}$ years. However, the final stellar mass is determined by 
processes such as radiative feedback from the protostar. 
We treat the evolution of a protostar as a sequence of a growing
hydrostatic core with an accreting envelope.  
The ordinary stellar
structure equations are applied to the hydrostatic core.  
The structure of the accreting envelope is calculated under the assumption
that the flow is steady for a given mass accretion rate.

Figure~3 shows the resulting evolution of the protostar.
After a transient phase and an adiabatic growth phase
at $M_{*} < 10 M_{\odot}$, the protostar enters the
Kelvin-Helmholtz phase and contracts by radiating its thermal energy.
When the central temperature reaches $10^{8}$K, 
hydrogen burning by the CNO cycle begins
with a slight amount of carbon synthesized by helium burning. 
This phase is marked by a solid circle in the figure. 
The energy generation by hydrogen burning halts
contraction when the mass is $35 M_{\odot}$ and its radius is 
$\sim 2.8$ solar radii. Soon after, the star reaches the 
zero-age main sequence (ZAMS). The protostar relaxes to a ZAMS star 
within about $10^5$ years
from the birth of the protostellar seed.
Accretion is not halted by radiation from
the protostar to the end of our calculation.

It is important to point out that the mass of the parent cloud from 
which the star formed is $M_{\rm cloud}\sim 40 M_{\odot}$. 
The final stellar mass is likely limited by the mass
of the gravitationally unstable parent cloud. 
We thus argue that primordial
stars formed from an ionized gas are massive, with a characteristic
mass of several tens of solar masses, allowing overall uncertainties
in the accretion physics and also the dependence of the minimum 
gas temperature on redshift (see next section).
They are smaller than the first stars formed from a neutral gas,
but are not low-mass objects as suggested by earlier
studies.

\begin{inlinefigure}
\resizebox{9.3cm}{!}{\includegraphics{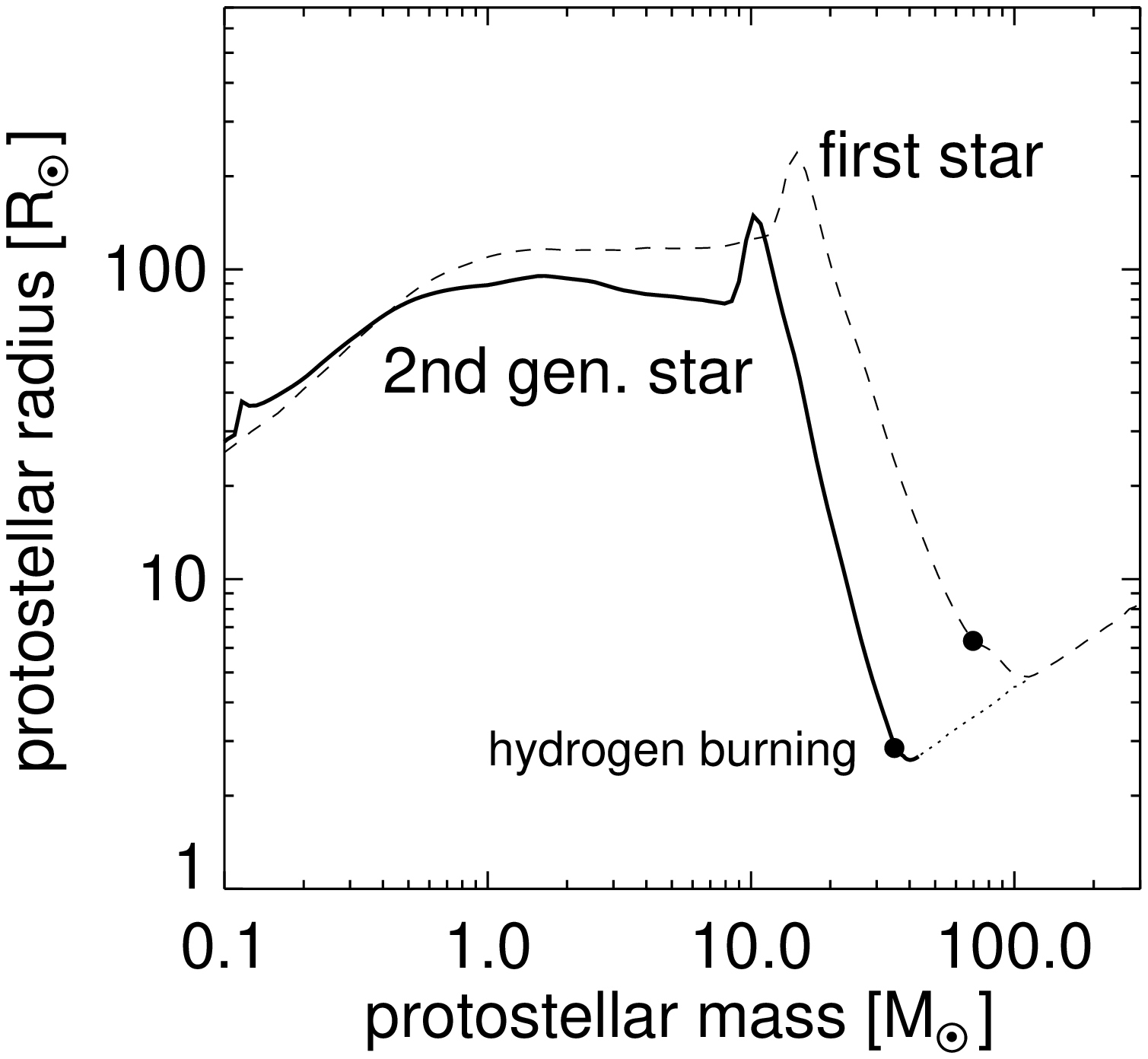}}
\caption{Evolution of the proto-stellar radius and the mass
(solid line). 
The solid circle marks the time when
efficient hydrogen burning begins. 
The dotted line shows the mass (and radius) growth
which is calculated under the assumption that a larger amount of gas
than the parent cloud can be accreted.
For reference we also show the result from
Y06 for a first generation star that forms
in a neutral gas cloud. 
\label{fig:proto}}
\end{inlinefigure}

\section{Discussion}
Recent theoretical studies of the formation of the first stars
indicate that they
were likely rather massive, with a characteristic mass greater than
a hundred solar masses (Abel et al. 2002; Omukai \& Palla 2003; 
Y06; Gao et al. 2007). 
We have explicitly shown, for the first time, that primordial stars 
formed from a reionized gas have a smaller mass ($\sim 40 M_{\odot}$)
than these first stars. The low temperature of the gas cloud owes to HD cooling,
and the correspondingly small collapse mass and low accretion rate 
lead to the smaller final stellar mass. We note that various proto-stellar feedback
effects such as mass outflow, which we do not take into account,
could briefly shut off accretion (Machida et al. 2006),
and then the final mass would be even {\it lower}.

Intriguingly, the elemental abundance patterns of hyper metal-poor stars 
discovered by Christlieb et al. (2002) and Frebel et al. (2005) 
indicate that early metal-enrichment was caused by supernova explosions
with a progenitor mass of $\sim 25 M_{\odot}$ (Umeda \& Nomoto 2003; Iwamoto et al. 2005).
Massive primordial stars formed in reionized regions, 
such as those studied here,
may be responsible for supernova explosions in the early Universe
(Tumlinson et al. 2004; Greif \& Bromm 2006).

We have shown that the minimum temperature of the gas cloud is
limited by the CMB, and thus the thermal evolution is
slightly different if the gas cloud collapses late 
at a lower redshift. 
We ran simulations varying the collapse redshift
(in practice, the CMB temperature) to examine this effect. 
The simulations show that the mass scale for the first runaway collapse 
becomes smaller, down to $M_{\rm cloud}\sim 10 M_{\odot}$, owing to the decrease
in the minimum gas temperature. 
While this might seem to suggest a lower characteristic mass for 
primordial stars formed at lower redshifts, the effect of photo-dissociating radiation 
in the far-ultra violet (FUV) needs to be considered carefully in this context,
because a very weak FUV radiation field is enough to destroy the H$_2$
and HD molecules in a
primordial gas cloud.

We carried out one-zone calculations of a collapsing primordial
gas as in Nagakura \& Omukai (2005), but with photodissociation of 
H$_{2}$, HD, and H$^{-}$.
The temperature evolution for primordial clouds with different FUV strength 
is shown in Figure 4.  We identify a critical flux
of $G_0=10^{-2}$ in terms of the Habing parameter (Habing 1968), 
or equivalently $3\times 10^{-22} {\rm erg}\;{\rm s}^{-1} {\rm cm}^{-2} {\rm Hz}^{-1}{\rm str}^{-1} $, 
above which the formation of (and hence cooling by) H$_2$ molecules is
limited by photo-dissociation, and the gas cloud starts 
contracting gravitationally
before its temperature is lowered to $< 100$ K by HD cooling.
The critical intensity is likely to be
reached before the completion of global reionization
(e.g., Haiman et al. 1997; Yoshida et al. 2003).
It is also expected that FUV radiation from a nearby source
exceeds this limit.
For example, a massive star with a few tens solar-masses emits 
photons in the Lyman-Werner bands at a rate of 
$L_{\rm LW}\sim 10^{24} {\rm erg}\; {\rm s}^{-1} {\rm Hz}^{-1}$.
For a typical physical condition in an early relic \HII region
(see Y07), HD formation will be suppressed if another massive star
is located within a radius of 
$r_{\rm diss.}=80\; (n_{\rm H_{2}}/10^{-3}{\rm cm^{-3}})^{-3/11}$ pc
even if we account for gas self-shielding.
The number of stars formed by HD cooling 
is likely limited within a relic \HII region, 
and thus the formation efficiency of these stars appears to be low.

It is also worth discussing the effect of metallicity.
Atomic metal line cooling at low densities 
will not significantly affect the evolution of the collapsing gas clouds,
because $T_{\rm CMB}$ determines the minimum gas temperature
even in the presence of metal cooling. 
On the other hand, the existence of dust may play an important role 
in significantly changing the characteristic mass of collapsing
gas clouds. Schneider et al. (2006) show that thermal dust
emission in such clouds can result in low temperatures at high gas densities, 
lowering the collapse mass scale to below one solar mass.
Therefore our conclusion remains robust 
unless a significant amount of dust is present in the gas clouds.

Finally, there is an interesting possibility of observing the death of 
massive primordial stars as gamma-ray bursts. Tominaga et al. (2007) 
recently suggested that primordial stars with mass $\sim 40M_{\odot}$ 
trigger long-duration
gamma-ray bursts and faint supernovae, such as those observed in the
nearby Universe (e.g., Fynbo et al. 2006). Future observations of 
bright gamma-ray bursts 
and the absorption lines in their radiation spectra will provide 
invaluable information on the interstellar medium
at high redshifts (Inoue, Omukai, \& Ciardi 2005).

\begin{inlinefigure}
\resizebox{9.3cm}{!}{\includegraphics{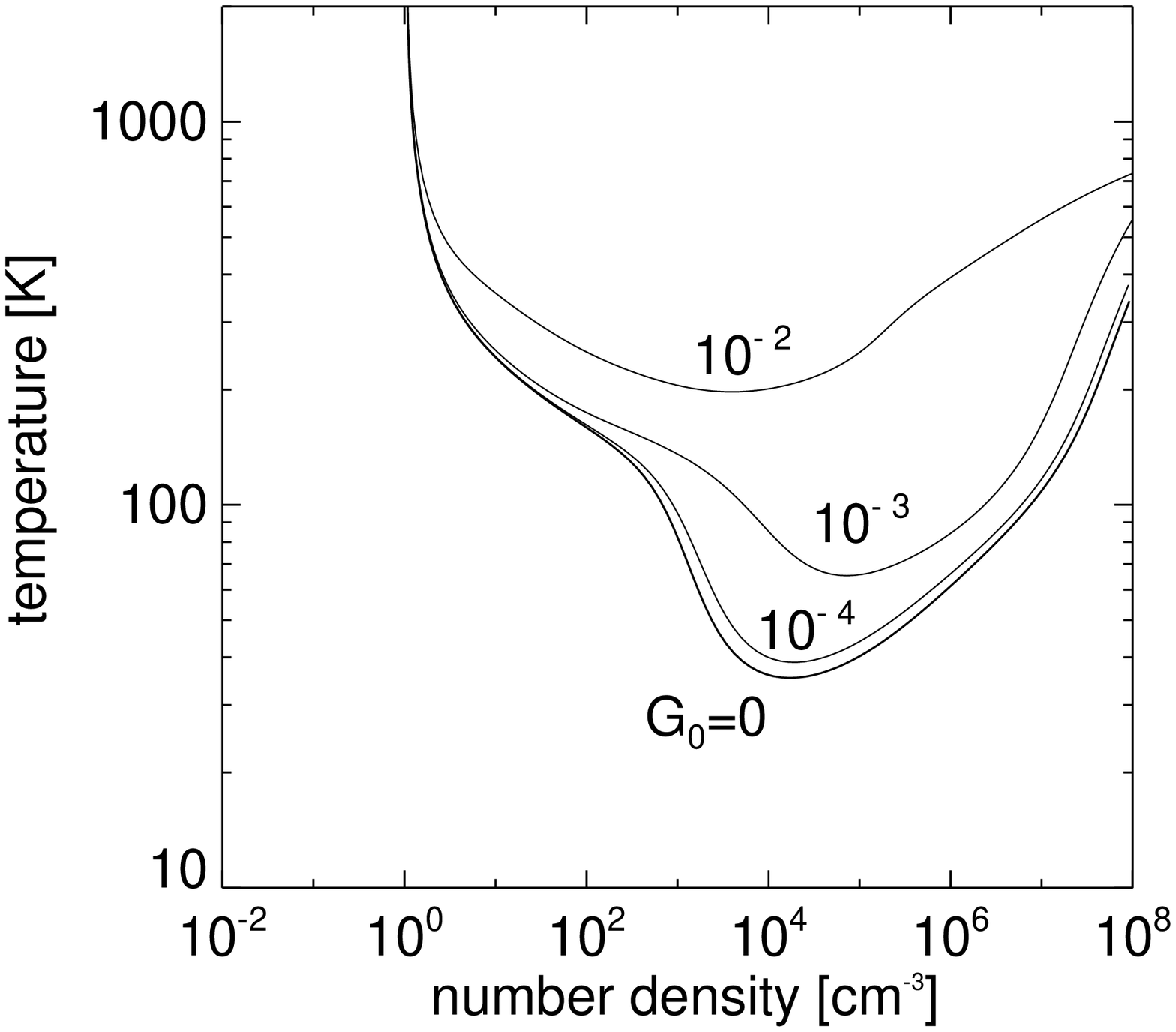}}
\caption{Temperature evolution of an initially ionized 
primordial gas with FUV radiation. 
The FUV radiation flux is parametrized by the Habing 
parameter $G_{0}$, which corresponds to the mean intensity of
$\sim 3 \times 10^{-20}G_{0} {\rm erg}\;{\rm s}^{-1} {\rm cm}^{-2} 
{\rm str}^{-1} {\rm Hz}^{-1}$ in the LW wavelength range. 
The initial number density, temperature and ionization degree are 
$1{\rm cm^{-3}}$, 7000K, and 0.1, respectively. 
\label{fig:UV}}
\end{inlinefigure}

We thank Ken Nomoto for discussions on
supernovae in the early universe. The work is supported in part by the Grants-in-Aid 
for Young Scientists 17684008 (NY) and 18740112 (TK). 
by the Ministry of Education, Culture, Science and 
Technology of Japan (NY). NY is grateful for financial
support from The Mitsubishi Foundation.


\begin{thebibliography}{24}

\bibitem{ABN} 
Abel, T., Bryan, G.~L., \& Norman, M. L. 2002, 
Science, 295, 93

\bibitem{Br02} 
Bromm, V., Coppi, P. S., \& Larson, R. B. 
2002, 
ApJ,
564, 23

\bibitem{Chris}
Christlieb, N. et al.
2002, Nature, 419, 904

\bibitem{couch}
Couchman, H.~M.~P. \& Rees, M.~J.,
1986, 
MNRAS, 221, 53 


\bibitem{self}
Draine, B. T., \& Bertoldi, F.,
1996, 
ApJ, 468, 269

\bibitem{fan}
Fan, X., et al. 
2006,
AJ, 132, 117

\bibitem{anna}
Frebel, A., Wako, A., Christlieb, N., Ando, H., Asplund, M.,
Barklem, P.-S., et al., 
2005,
Nature, 434, 871

\bibitem{furl2004}
Furlanetto, S., Zaldarriaga, M., \& Hernquist, L., 2004,
ApJ, 613, 1

\bibitem{furl2006}
Furlanetto, S., McQuinn, M., \& Hernquist, L., 2006,
MNRAS, 365, 115

\bibitem{fynbo}
Fynbo, J.~P.~U. et al.
2006,
Nature, 444, 1047

\bibitem{gao06}
Gao, L., Yoshida, N., Abel, T., Frenk, C.~S., Jenkins, A., Springel, V.,
2007, 
MNRAS, 378, 449

\bibitem{gnedin} 
Gnedin, N. \& Ostriker, J. P.,
1997, 
ApJ, 486, 581



\bibitem{greif} 
Greif, T. H. \&  Bromm, V.,
2006,
MNRAS, 373, 128


\bibitem{habing}
Habing, H.~J.,
1968, BAN, 19, 421

\bibitem{haiman1997}
Haiman, Z., Rees, M.J., \& Loeb, A., 1997, ApJ, 476, 458


\bibitem{inoue}
Inoue, S., Omukai, K., Ciardi, B., 
2005, astro-ph/0502218


\bibitem{iwamoto}
Iwamoto, N., Umeda, H., Tominaga, N., Nomoto, K.,
\& Maeda, K., 2005, Science, 309, 451


\bibitem{john}
Johnson, J.~L., \& Bromm, V.,
2006,
MNRAS, 366, 247

\bibitem{Kashi}
Kashikawa, N., et al.
2006,
ApJ, 648, 7

\bibitem{lenz91}
Lenzuni, P., Chernoff, D.F, \& Salpeter, E.E.,
1991, ApJS, 76, 759

\bibitem{machida}
Machida, M. N., Omukai, K., Matsumoto, T., Inutsuka, S.,
2006,
ApJL, 647, 1


\bibitem{mcquinn06} 
McQuinn, M., Lidz, A., Zahn, O., Dutta, S., 
Hernquist, L., Zaldarriaga, M.,
2006, MNRAS, submitted [astro-ph/0610094]

\bibitem{Miralda} 
Miralda-Escud\'e, J., Haehnelt, M., \& Rees, M.,
2000,
ApJ, 530, 1

\bibitem{NO05}
Nagakura, T. \& Omukai, K.,
2005,
MNRAS, 364, 1378

\bibitem{NM02} 
Nakamura, F. \& Umemura, M.,
2002,
ApJ, 569, 549

\bibitem{on2} 
Omukai, K. \& Nishi, R.,
1998, 
ApJ,
508, 141

\bibitem{OP03}
Omukai, K. \& Palla, F.,
2003, 
ApJ,  589, 677

\bibitem{OY03}
Omukai, K. \& Yoshii, Y.,
2003, 
ApJ, 599, 746

\bibitem{LowZ}
Omukai, K., Tsuribe, T., Schneider, R., \& Ferrara, A.,
2005, 
ApJ,
626, 627

\bibitem{wmap3p}
Page, L., et al.
2007,
ApJS, in press

\bibitem{sok03}
Sokasian, A., Abel,. T., Hernquist, L., \& Springel, V.,
2003, MNRAS, 344, 607

\bibitem{sok04}
Sokasian, A., Yoshida, N., Abel,. T., Hernquist, L., \& Springel, V.,
2004, MNRAS, 350, 47

\bibitem{wmap3}
Spergel, D.~N., et al.
2007, ApJS, in press

\bibitem{stahler80a}
Stahler, S.W., Shu, F.H., \& Taam, R.E., 1980,
ApJ, 241, 637

\bibitem{stahler86}
Stahler, S.W., Palla, F., \& Salpeter, E.E., 1986,
ApJ, 302, 590

\bibitem{tomi} 
Tominaga, N., Maeda, K., Umeda, H., Nomoto, K., Tanaka, M.,
Iwamoto, N., Suzuki, T., Mazzali, P.~A.,  
2007,
ApJ, 657, L77


\bibitem{tum04} 
Tumlinson, J., Venkatesan, A., \& Shull, J.~M.,
2004,
ApJ, 612, 602


\bibitem{grb06} 
Totani, T. et al. 
2006,
PASJ, 58, 485

\bibitem{uehara} 
Uehara, H. \& Inutsuka, S.,
2000,
ApJ, 531, 91

\bibitem{umeda03} 
Umeda, H. \& Nomoto, K.,
2003,
Nature, 422, 871

\bibitem{y03} 
Yoshida, N., Abel, T., Hernquist, L. \& Sugiyama, N., 
2003, 
ApJ, 592, 645

\bibitem{y06} 
Yoshida, N., Omukai, K., Hernquist, L. \& Abel, T., 
2006, ApJ, 652, 6 [Y06]

\bibitem{y07} 
Yoshida, N., Oh, S.-P., Kitayama, T., \& Hernquist, L., 
2007, ApJ, 663, 687 [Y07]

\bibitem{zahn07} 
Zahn, O., Lidz, A., McQuinn, M., Dutta, S., 
Hernquist, L., Zaldarriaga, M., Furlanetto, S.~R., 
2007,
ApJ, 654, 12

\end{thebibliography}
\end{document}